\documentclass[twoside,12pt]{article}

\usepackage{amsmath}
\usepackage{amssymb}
\usepackage{graphicx}

\def\Journal#1#2#3#4{{#1} {\bf #2} (#4) #3 }

\def\PLB{{\em Phys. Lett.} B}

\def\PRL{\em Phys. Rev. Lett.}

\def\PRE{{\em Phys. Rev.} E}
\def\PRC{{\em Phys. Rev.} C}
\def\PRB{{\em Phys. Rev.} B}

\def\RMP{{\em Rev. Mod. Phys.}}

\newcommand{\be}{\begin{equation}}
\newcommand{\ee}{\end{equation}}
\newcommand{\bea}{\begin{eqnarray}}
\newcommand{\eea}{\end{eqnarray}}

\newcommand{\conj}[1]{\ensuremath{#1^{*}}}

\newcommand{\onebody}[3]{\ensuremath{\langle{#1}|{#2}|{#3}\rangle}}

\newcommand{\bvec}[1]{\ensuremath{\boldsymbol{#1}}}

\newcommand{\tgenm}[1]{\ensuremath{\boldsymbol{\mathcal{#1}}}}

\newcommand{\matnot}[1]{\ensuremath{\mathbf{#1}}}

\newcommand{\op}[1]{\ensuremath{\hat{#1}}}

\newcommand{\lat}[1]{\ensuremath{\mathfrak{#1}}}

\newcommand{\Hamm}[0]{\ensuremath{\mathcal{H}}}

\begin{document}

\title{ \vspace{1cm} Towards a quantal dynamical simulation of the neutron-star crust}
\author{Klaas Vantournhout $^{1}$, Thomas\ Neff $^1$, Hans Feldmeier $^1$,\\
Natalie Jachowicz $^2$, Jan Ryckebusch $^2$\\
\\
$^1$GSI Helmholtzzentrum f\"{u}r Schwerionenforschung GmbH, Darmstadt, Germany\\
$^2$Ghent University, Department of Physics and Astronomy, Gent, Belgium}
\maketitle
\begin{abstract}
We present a novel method to study the dynamics of bulk fermion systems such as the neutron-star crust. By introducing periodic boundary conditions into Fermionic Molecular Dynamics, it becomes possible to examine the long-range many-body correlations induced by antisymmetrisation in bulk fermion systems. The presented technique treats the spins and the fermionic nature of the nucleons explicitly and permits investigating the dynamics of the system. Despite the increased complexity related to the periodic boundary conditions, the proposed formalism remains computationally feasible.
\end{abstract}

\section{Introduction}
Among the myriad of phenomena in the universe, a neutron star is one of the extremely interesting objects. As an ``oversized nucleus'', it is one of the most dense and compact objects known. The densities of the inner part of a neutron star are expected to be as large as five to ten times the normal nuclear saturation density $n_0$. This leads to intriguing speculations about the composition of the neutron-star core \cite{haensel2007}. The densities of the outer layers of a neutron star are on the average below $n_0$. At these densities, the neutron-star crust consists of protons, neutrons and electrons, organising themselves as a result of the short-range nuclear attraction and the long-range Coulomb repulsion.\\

In the outer crust of a neutron star, i.e.~the lower-density regions, nucleons bind in nuclei and are placed on a Coulomb lattice, embedded in an electron gas. These nuclei grow with increasing density. When entering the inner crust, the nuclei surpass the neutron-drip densities and immerse the Coulomb lattice in a neutron gas. Finally, in the densest layers of the crust ($n_0/3$ to $n_0/2$ \cite{haensel2007}), the spherical nuclear clusters outgrow the unit cell of the Coulomb lattice, transforming the crystallised crust into uniform core matter. In this crust-core interface, also known as the mantel of the neutron star, the system balances on a subtle interplay between the nuclear surface energies and the Coulomb energy of the neutron-proton-electron system. This results in a multitude of competing quasi-ground states, having sometimes quite different matter distributions, where the system can cool down to. These complex-shaped structures are dubbed ``nuclear pastas'' \cite{ravenhall1983, sonoda2008, chamel2008}.\\

The crust-core interface of a neutron star may represent a sizable fraction of the crustal mass \cite{lorenz1993}. After all, it represents the densest regions of the crust. Hence, to apprehend the crust is an important key to understanding many astrophysical phenomena related to neutron stars. Furthermore, matter at sub-saturation densities, including nuclear pastas, can contribute 10-20\% of the mass in the later stages of a collapsing stellar core \cite{sonoda2007}, making them also important for supernovae physics. As such, a sound description of the neutron-star crust, or nuclear matter at sub-saturation densities in general, is needed to understand various astrophysical observations \cite{chamel2008}.\\

The nuclear pasta phases have been investigated with a broad spectrum of computational techniques ranging from the liquid drop model over Hartree Fock to Monte Carlo and molecular dynamics \cite{ravenhall1983, oyamatsu1984, watanabe2009, nakazato2009, newton2009, avancini2010}. From all these, molecular dynamics is one of the few that is not biased with regard to the geometry of the nuclear clusters. Moreover, thanks to the dynamical aspect, one can go beyond the study of ground-state structures. The latter is most interesting for nuclear pastas as, with a preponderance of low-energy excitation levels, they are susceptible to low-energy dynamics stemming from external probes or temperature changes.\\

\section{Fermionic Molecular Dynamics}

Molecular dynamics techniques have already been successfully deployed to study various properties of nuclear pastas and the neutron-star crust. Classical Molecular Dynamics (CMD) provided a way to study the effect of large density fluctuations on neutrino opacities and to evaluate the breaking strain of the crust \cite{horowitz2004,horowitz2009}. A more advanced model, named Quantum Molecular Dynamics (QMD), was used to identify the various pasta structures and to determine the possible transitions between all those structures \cite{watanabe2009}.\\

In CMD, the nucleons are represented by points in classical phase space. The QMD model, on the other hand, uses localised states and mimics the Pauli principle through a potential. Nevertheless, QMD can still be interpreted as CMD with the classical Hamilton equations of motion\,:
\begin{equation}
\dot{\bvec q_i} = \frac{\partial H}{\partial \bvec p_i},\qquad
\dot{\bvec p_i} = -\frac{\partial H}{\partial \bvec q_i}.
\end{equation}

When studying phenomena at length scales smaller than the de Broglie wavelength of the system's constituents, CMD breaks down. Under those conditions, quantum effects stemming from the wave character of the nucleons predominate, and statistical many-body correlations related to the system's fermionic content play a fundamental role.  In QMD, the effect of fermionic correlations is mimicked by means of a phenomenological Pauli potential. Although this method is quite successful in calculating energies, it fails at reproducing the single-particle occupations in momentum space for the free Fermi gas \cite{dorso1987,dorso1988}.  Furthermore, in QMD, the spin and the width of the localised states are assumed to be time independent.\\

The shortcomings of QMD for fermion systems are remedied in Antisymmetric Molecular Dynamics (AMD) \cite{ono2004} or the more general Fermionic Molecular Dynamics (FMD) \cite{feldmeier2000}. In these techniques, the many-body state is introduced as an antisymmetrised product, i.e.~Slater determinant, of localised single-particle states $|\phi_p(\bvec z_p)\rangle$, denoted as:
\begin{equation}\label{eq:trialstate}
|\Phi\rangle = \op{A}|\phi_1(\bvec z_1)\rangle\otimes\cdots\otimes|\phi_A(\bvec z_A)\rangle.
\end{equation}
Here, $\bvec z_p$ represents the complex time-dependent parametrisation of the single-particle state $|\phi_p\rangle$. The time evolution of these parameters is determined by the time-dependent variational principle \cite{kramer1981} which leads to the generalised Hamilton equations of motion\,:
\begin{equation}\label{eom}
i\sum_{q=1}^A\matnot{C}_{pq}\dot{\bvec z}_q^{\phantom\star} =
\frac{\partial\Hamm}{\partial\conj{\bvec z}_p},\qquad \matnot C_{pq} = \left(\frac{\partial^2\matnot n_{pq}}{\partial\conj{\bvec z}_p\partial
  \bvec z_q}-\sum_{rs=1}^A\frac{\partial\matnot
      n_{pr}}{\partial\conj{\bvec z}_p}\cdot\matnot o_{rs}\cdot\frac{\partial\matnot
  n_{sq}}{\partial\bvec z_q}\right)\cdot\matnot o_{qp}.
\end{equation}
In this equation, $\matnot C$ is the metric of the parameter space of $|\Phi\rangle$ and $\Hamm = \onebody{\Phi}{\Hamm}{\Phi}/\langle\Phi|\Phi\rangle$ is the Hamiltonian expectation value. The matrix $\matnot o$ represents the inverse of the overlap matrix $\matnot n$ with $\matnot n_{pq} = \langle \phi_p|\phi_q\rangle$.\\

The time-evolved many-body state is used to study the properties of the Fermi system by evaluating various $N$-body operators. The expectation values of one- and two-body operators are calculated as
\begin{equation}\label{eq:finite:operators}
\mathcal{B}_I =
\sum_{pq=1}^A\onebody{\phi_p}{\op{B}_{I}}{\phi_q}\matnot{o}_{qp},\qquad
\mathcal{B}_{II}=
\frac{1}{2}\sum_{pqrs=1}^A \onebody{\phi_p\phi_r}{\op{B}_{II}}{\phi_q\phi_s}
(\matnot{o}_{qp}\matnot{o}_{sr} -
\matnot{o}_{qr}\matnot{o}_{sp}).
\end{equation}
Because of the determinant structure of $|\Phi\rangle$, the equations relevant to FMD can be written in matrix notations using the inverse overlap matrix $\matnot o$. Thereby, the matrices $\matnot n$ and $\matnot o$ play a fundamental role as they carry all information about the fermion statistics of the system under study. As their eigenvalues can cover many orders of magnitudes, it is paramount to calculate these matrices as accurately as possible, i.e.~analytically.\\

Both AMD and FMD enjoy successes in the study of heavy-ion collisions, ground states of nuclei and astrophysical reaction calculations \cite{ono2004, neff2008, neff2010}. However, due to the long-range character of the Pauli correlations, evaluating bulk fermion matter becomes a tedious task. The dimension of the matrices usually impedes simulations of a large number of particles, however desirable these may be for bulk fermion systems such as crustal matter. 

\section{Bulk fermionic molecular dynamics}

When investigating the properties of bulk matter by means of large simulation volumes, the evaluation of expectation values becomes cumbersome. Moreover, surface effects may influence the results when a large fraction of the constituents lies on the surface of the simulation volume. A time-honoured method  is to introduce a periodic structure in the simulation.  When studying bulk matter using a periodic structure, it is crucial to make sure that the studied properties of the small but infinitely repeated periodic system and the macroscopic system which it represents, are the same.  As long as the correlation volume of the interactions does not exceed the simulation volume, the imposed periodicity works fine.  However, serious problems arise in the presence of long-range correlations as e.g.~induced by Coulomb interactions or by the Pauli exclusion principle.\\

When studying bulk fermion systems in a mean-field approach using Slater determinants, the single-particle states are generally delocalised single-particle states fulfilling certain boundary conditions. These could be ``periodic boundary conditions'', ``Bloch conditions'' or ``twist-averaged boundary conditions'' \cite{ashcroft1976,lin2001}. Spatial localisations are, however, common in the neutron-star crust. To describe these with delocalised states, a large configuration space is required. The localised states, as used in FMD, would circumvent this problem but are difficult to use for bulk inhomogeneous matter.\\

As the Pauli exclusion principle introduces long-range many-body correlations, it poses a major challenge to the study of bulk fermion systems with non-orthogonal localised states. It is indispensable to fully antisymmetrise the many-body wave function representing the system. To simulate bulk matter with localised states, we propose to make the spatial positioning of the single-particle states periodic. A number of particles are placed in unit cells on a lattice \lat B, that is tessellating space perfectly.  Each cell, containing $A$ particles, can be identified by a lattice vector $\bvec R=n_1\bvec a_1 + n_2\bvec a_2 + n_3\bvec a_3$ with integer $n_j$. The trial state of the system is thus, with $\op{T}(\bvec R)$ the translation operator over $\bvec R$,
\begin{equation}\label{eq:tapbc:W}
  |\Phi_\infty\rangle = \op{A}\bigotimes_{\bvec R\in\lat B} \op{T}(\bvec R)
  \left\{|\phi_1\rangle\otimes\cdots\otimes|\phi_A\rangle\right\}.
\end{equation}

For $|\Phi_\infty\rangle$, the overlap matrix ${\matnot n}$ and its inverse ${\matnot o}$ have infinite dimensions. Nonetheless, they exhibit a peculiar nested block-Toeplitz structure which can be exploited \cite{vantournhout2010}.  As a result of the translational invariance of the system, each $A\times A$ block of the overlap matrix can be identified unambiguously with the lattice vector $\bvec R$ that connects the two cells of the bra and ket states. The blocks can be evaluated as $\matnot n_{pq,\bvec R} = \langle \phi_p|\op{T}(-\bvec R)|\phi_q\rangle$. Although $\matnot n$ and $\matnot o$ are of infinite dimension, it is possible to obtain the inverse overlap matrix through the following scheme
\begin{equation}
\tgenm N(\bvec k) = \sum_{\bvec R\in\lat B} \matnot n_{\bvec R}\,
  e^{-i\bvec k\cdot\bvec R},\qquad \tgenm O(\bvec k) = \tgenm
  N(\bvec k)^{-1}, \qquad\matnot o_{\bvec R} = \frac{1}{V_{BZ}}\int_{BZ}\tgenm O(\bvec
  k)\,e^{i\bvec k\cdot\bvec R}\,d^3\bvec k,
\end{equation}
where $BZ$ represents the first Brillouin zone of the lattice $\lat B$ and $\bvec k$ is a vector in this volume \cite{vantournhout2010}.\\

The strength of the proposed formalism reveals itself upon evaluating the operators in reciprocal space.  Normally, the expectation values of the infinite fermion system would be calculated by means of Eqs.~\eqref{eq:finite:operators} resulting in infinite sums over the block structure. These sums, however, translate into integrals over the first Brillouin zone which are computationally straightforward to evaluate. The expectation value of a one-body operator per unit-cell volume can then be computed as
\begin{equation}\label{eq:pbc:operators:bloch}
    \mathcal{B}_{\rho,I} = \frac{1}{V_{BZ}}\int_{BZ} \sum_{pq=1}^A\tgenm
    B_{I,pq}(\bvec k)\tgenm O_{qp}(\bvec k)\,d^3\bvec k ,\qquad
    \tgenm B_{I,pq}(\bvec k) = \sum_{\bvec R\in\lat B}
    \onebody{\phi_p}{\op{B}_{I}\op{T}(\bvec R)}{\phi_q}\,
    e^{i\bvec k\cdot\bvec R}.
\end{equation}
Here we assume that the operator $\op{B}_I$ commutes with the translation operator $\op{T}(\bvec R)$. The two-body operator and the metric have analogue structures \cite{vantournhout2010}.\\

From a computational perspective, a proper choice of the localised single-particle states helps speeding-up the computations. Gaussian wave packets come with the interesting feature that most matrix-elements can be evaluated analytically.  Furthermore, the computational order for the calculation of expectation values is the same for bulk fermion systems as for finite-sized systems ($N^4$ for two-body interactions), even though the former represent a system with an infinite number of particles. On the other hand, the equations for the bulk systems involve an extra integration over the Brillouin zone of the imposed lattice, increasing the computational effort. However, it has been shown that integrals over the Brillouin zone can be performed using only a limited number of specific $\bvec k$-points \cite{monkhorst1976}.

\section{Results}

\begin{figure}[ht!]
\begin{center}
\includegraphics[width=0.85\textwidth]{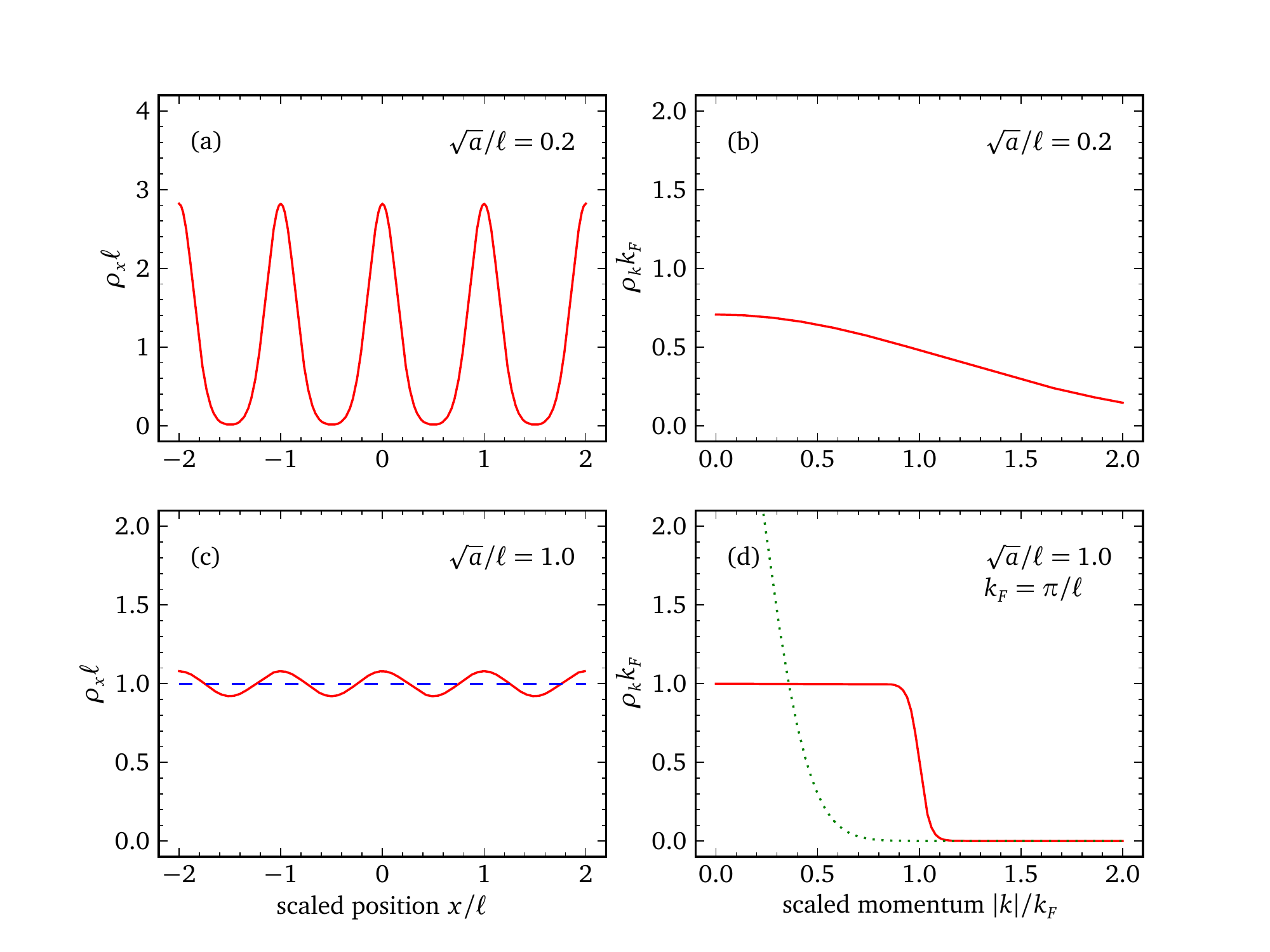}
\caption{Spatial density (panels (a) and (c)) and momentum density (panels (b) and (d)) of equidistant Gaussian wave packets with one-dimensional periodicity. One Gaussian with variance $a$ per unit cell of size $\ell$. For small overlap (top) the system behaves like one of distinguishable particles while for large overlap (bottom) the fermionic behaviour becomes manifest. For comparison, the spatial distribution of a uniform Fermi gas ($\sqrt{a}/\ell\rightarrow\infty$) is presented in panel (c) (dashed line) and the momentum distribution of distinguishable particles is presented in panel (d) (dotted line).}\label{fig:1ddist}
\end{center}
\end{figure}

In the following, we will show that the proposed technique reproduces the features intrinsic to the fermionic behaviour of the system. The unnormalised single-particle states are Gaussian wave packets of the form $\langle\bvec x|a\bvec b\rangle = \exp\{-(\bvec x - \bvec b)^2/(2a)\}$ where the complex vector $\bvec b$ represents the mean position in phase space and $a$ is a complex parameter connected with the width of the wave packet. In Ref.~\cite{feldmeier2000} it was shown that for a hundred periodically positioned wave packets, Eqs.~\eqref{eq:finite:operators} reproduce the momentum and spatial densities intrinsic to one-dimensional Fermi systems. In Fig.~\ref{fig:1ddist}, we show that our method also reproduces this result by placing a single particle in a unit cell of size $\ell$. It is clearly visible that with increasing $\sqrt{a}/\ell$ and thus with increasing overlap of neighbouring states, hence increasing influence of the antisymmetrisation, the fermionic behaviour becomes apparent. For $\sqrt{a}/\ell=1$, the momentum distribution does not show a sharp cutoff near the Fermi momentum because the periodic set of Gaussian wave packets does not constitute a complete set. For $\sqrt{a}/\ell\rightarrow\infty$, the antisymmetry operator projects the Gaussian wave packets on plane waves and a sharp cutoff is reached.
Even though this one-dimensional example clearly hints at free-fermion behaviour, it has to be assured that fermion-like properties of the simulations are not an artifact of the symmetry imposed by the lattice. In fact, it can be shown that in more dimensions the momentum distribution evolves towards the first Brillouin zone of the lattice \cite{vantournhout2010}. This effect can be expected as, with increasing overlap, the wave packets seem uniform to the unit cell and the system behaves as plane waves in a periodic system.\\

\begin{figure}[ht!]
\begin{center}
\includegraphics[width=0.85\textwidth]{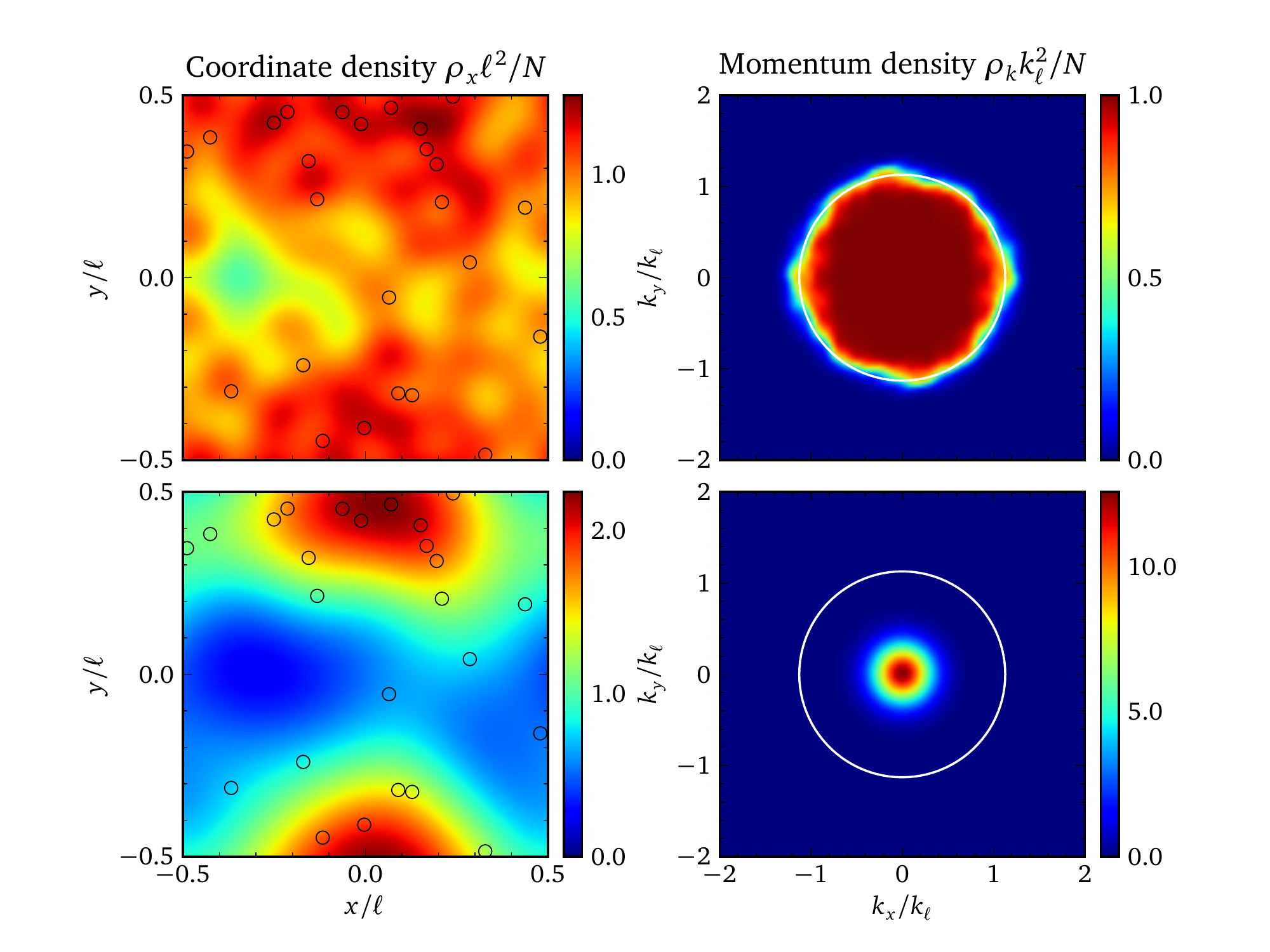}
\caption{Spatial and momentum distribution of $N=25$ Gaussian wave packets with periodic boundary conditions. The wave packets are randomly placed in the unit cell of a square lattice with unit length $\ell$ and have a variance given by $\sqrt{a}=0.2\ell$. The mean momentum of the individual wave packets is set to zero. The densities are normalised and coordinates are expressed in units of $\ell$ or $k_\ell = \sqrt{N}\pi/\ell$. The upper two panels depict the antisymmetrised case.  The lower two panels represent distinguishable particles. The black circles show the positions of the centroids of the wave packets. The white circle represents the Fermi ``sphere'' of a system of free particles with the same mean density, $k_F = \sqrt{4\pi N}/\ell$.} \label{fig:2dsqrdist}
\end{center}
\end{figure}

As we introduced periodic boundary conditions to eliminate surface effects, a successful investigation of bulk matter obviously requires that influences of the geometry of the chosen boundary conditions are negligible. In Fig.~\ref{fig:2dsqrdist} we show that this is achieved for a situation where the 25 single-particle states are randomly distributed in the cell. The figure shows that the spatial and momentum distribution of the system reflects that of a free fermion system. The perturbations on the Fermi-sphere in momentum space, as well as those on the uniform distribution in coordinate space can be understood as a result of the local clustering. Hence, a simulation with the proposed periodic structure reproduces the intrinsic bulk fermion behaviour, independent of the imposed periodic structure. As was indicated earlier, this result was hard to reach using QMD. For comparison, the bottom panels of Fig.~\ref{fig:2dsqrdist} display the densities for distinguishable particles occupying the same wave packets. Both the spatial and momentum distributions are quite different from the indistinguishable fermion case. It clearly shows that antisymmetry induces Fermi motion and redistributes density in coordinate and momentum space. This long-range many-body correlation should not be ignored. 

\section{Conclusion}

In summary, we have implemented periodic boundary conditions in the FMD formalism. This allows one to study infinitely extended non-uniform fermionic matter with full antisymmetrisation. The introduced spatial periodicity in the trial-state leads to a Toeplitz structure for the FMD matrices, a feature that can be exploited to reduce the numerical cost of the simulation. The structure of the resulting equations resemble those of finite FMD done on a single unit cell, however with additional integrations in the reciprocal lattice space. This new formalism offers perspectives to compute the expectation values of various operators for bulk fermion matter in a computationally feasible fashion. Although the equations only address a finite number of particles $N$ in a unit cell, they keep track of the Fermi statistics of the infinite system. We have evaluated the validity of this description with a study of the spatial and momentum distribution of various lattice systems and showed that our simulations convincingly reproduce free fermion behaviour. This shows that full antisymmetrisation can be imposed in the bulk FMD formalism.\\

This work was supported by the Fund for Scientific Research Flanders (FWO), the Research Council of Ghent University and the Helmholtz International Center for FAIR (HIC for FAIR).

\end{document}